\documentclass[%
 reprint,
nofootinbib,
 amsmath,amssymb,
 aps,
pra,
]{revtex4-2}

\usepackage{revtex}
\usepackage{comandi_tesi}

\definecolor{applegreen}{rgb}{0.55, 0.71, 0.0}
\definecolor{cornellred}{rgb}{0.7, 0.11, 0.11}
\definecolor{darkolivegreen}{rgb}{0.33, 0.42, 0.18}
\definecolor{olivedrab}{rgb}{0.42, 0.56, 0.14}

\begin{document}
\title{Improving the speed of variational quantum algorithms\\
for quantum error correction}

\author{Fabio Zoratti}
\affiliation{Scuola Normale Superiore, I-56126 Pisa, Italy}
\author{Giacomo De Palma}
\affiliation{Department of Mathematics, University of Bologna, 40126 Bologna, Italy}
\author{Bobak Kiani}
\affiliation{Department of Electrical Engineering and Computer Science, MIT, 77 Massachusetts Avenue, Cambridge, MA 02139, USA}
\author{Quynh T. Nguyen}
\affiliation{Department of Electrical Engineering and Computer Science, MIT, 77 Massachusetts Avenue, Cambridge, MA 02139, USA}
\author{Milad Marvian}
\affiliation{Department of Electrical and Computer Engineering and Center for Quantum Information and Control, University of New Mexico, USA}
\author{Seth Lloyd}
\affiliation{\mbox{Department of Mechanical Engineering, MIT 77 Massachusetts Avenue, Cambridge, MA 02139, USA} \\Turing Inc., Cambridge, MA 02139, USA}
\author{Vittorio Giovannetti}
\affiliation{NEST, Scuola Normale Superiore and Istituto Nanoscienze-CNR, I-56126 Pisa, Italy}

\begin{abstract}
 We consider the problem of devising a suitable Quantum Error Correction (QEC) procedures for a generic quantum noise acting on a quantum circuit. In general, there is no analytic universal procedure to obtain the encoding and correction unitary gates, and the problem is even harder if the noise is unknown and has to be reconstructed. The existing procedures rely on Variational Quantum Algorithms (VQAs) and are very difficult to train since the size of the gradient of the cost function decays exponentially with the number of qubits. We address this problem using a cost function based on the Quantum Wasserstein distance of order 1 ($QW_1$). At variance with other quantum distances typically adopted in quantum information processing, $QW_1$
 lacks the unitary invariance property which makes it a suitable tool to avoid to get trapped in
 local minima.
 Focusing on a simple noise model for which an exact QEC solution is known and can be used as a theoretical benchmark, we run a series of numerical tests that show
how, guiding the VQA search through the $QW_1$, can indeed significantly increase both the probability of a successful training and the fidelity of the recovered state, with respect to the results one obtains when using conventional approaches.\end{abstract}

\maketitle

\section{Introduction}

Performing reliable computations on physically imperfect hardware is something that has become usual nowadays, given the current state of classical computers, which can produce perfect results without any software-side mitigation of the imperfections of the physical media where the computation happens. Error correction is based on the fact that these machines automatically perform, on the hardware side, procedures that allow errors to happen and to be fixed without any intervention from the end user. This kind of setting is even more crucial in a quantum scenario where the current noisy intermediate-scale quantum computers (NISQ) have a much larger error rate than their classical counterparts~\cite{nisq_review_2022}. Performing reliable computations with a trustworthy error correction procedure has direct implications not only in quantum computation~\cite{preskill_2012, Preskill_2018}, but potentially also in all the other sectors of quantum technology which indirectly relay on it~(e.g.~quantum communication or quantum key distribution~\cite{Gisin2002, Lo2014,Pirandola2020}).

In the typical Quantum Error Correction (QEC) scheme, the quantum information that has to be protected is stored in a subspace of a larger Hilbert space, using an \emph{encoding} procedure.
Stabilizer codes~\cite{gottesman1997stabilizer,knill2001_5qubits}, which are within the best analytical results in this field, are not universal because they are tailored for a generic noise acting on a small but unknown subset of qubits. Several attempts have already been made to create a numerical optimization procedure to find an error correction code for specific noise models~\cite{fletcher_2008, robust_qec_2008, soraya_2010, Chiani2020}, but these studies are not universal because they rely heavily on the type of noise on the specific quantum circuit and this is a problem because real quantum devices are not characterized by a single kind of quantum noise. Some attempts have been made to characterize the noise of the current and near-term devices~\cite{Koch_2007,PhysRevLett.114.010501}, but these methods will become very difficult to implement soon because classical computers are not able to simulate efficiently quantum circuits when the number of qubits increases. Near-term devices with approximately 50 qubits may already be intractable to simulate for supercomputers~\cite{Boixo_2018}.

If we define a figure of merit of the quality of the state after the action of the noise and its corresponding correction, the obvious choice for the kind of maximization algorithm is a Variational Quantum Algorithm (VQA)~\cite{cerezo_vqa2021}. These are hybrid algorithms that couple a quantum computer with a classical one. In this setting, usually, a parametric quantum circuit is applied to some reference state, some measurements are performed on the system, and the outcomes are given to the classical computer to perform a minimization procedure of a given cost function (from this point of view
the optimization procedure in a VQA can be seen as the training phase in machine learning). Some examples of this class of algorithms are the variational quantum eigensolver~\cite{vqe_review_2021} and the Quantum Approximate Optimization Algorithm~\cite{qaoa}. Proposals to use VQAs to address QEC problems are already present in literature~\cite{johnson2017qvector}. Unfortunately, VQAs usually suffer from the phenomenon of barren plateaus~\cite{barren_plateaus_2018,cerezo_barren_2021}, namely the gradient of the cost function decays exponentially with respect to the number of qubits of the system, leading to an untrainable model. The fundamental theoretical reason for such behavior has been
associated with the presence of {\it barren plateaus} which originate when the cost function of the problem
is global, i.e. mediated by a highly non-local operator~\cite{cerezo_barren_2021}.
 To avoid these effects we propose here to guide the VQA search using cost functions inspired to
 Quantum Wasserstein distance of order~$1$ (or $QW_1$ in brief) introduced in
 Ref.~\cite{gdp_wasserstein_order_1} as a quantum generalization of the Hamming distance \cite{hamming_distance} on the set of bit strings. As will detail in the following, at variance with more conventional
 quantum distances typically adopted in quantum information, $QW_1$
 is lacking a fundamental symmetry (unitary invariance) which makes it a suitable candidate to
 avoid the barren plateau problem. The rationale behind this is that
for unitarily invariant distances as the trace distance or the distances derived from the fidelity,
all the states of the computational basis are equally orthogonal and thus have all maximum distance one with respect to the other. The $QW_1$ functional instead measures how many qubits are different between the two states allowing the VQA gradient to be less flat in the regions that are not already very close to a local minimum.
While this special property of $QW_1$ has been already observed in other contexts, such as the study of
 quantum Generative Adversarial Networks presented in~\cite{gdp2021ghz,kim2022hamiltonian,herr2021anomaly,anschuetz2022beyond,coyle2022machine,chakrabarti2019quantum}, here we test its effectiveness in the identifying effective QEC procedures.
 For this purpose, we run a series of numerical tests which compare the performances of a VQA
 that adopts a conventional (i.e. unitary invariant) cost function, with that of a VQA which instead
 refers to $QW_1$-like distances. Our findings confirm that in the second case the effectiveness of the numerical optimization
 significantly increases both in terms of the probability of a successful training and in
 the fidelity of the recovered state.

The manuscript is organized as follows: in \cref{sec:W1} we present a concise, yet rather complete
review on the $QW_1$ distance for qubits;
in Sec.~\ref{sec:introQEC} we present some basic notions on conventional QEC procedures which allow us to set the notation and the
theoretical background; in Sec.~\ref{sec:general} we introduce our VQA discussing the different choices of cost functions that can be used in order to guide it; in Sec.~\ref{sec:res} we present our numerical results where comparing the performances of the VQA implemented with different types of cost functions. Conclusions are given in Sec.~\ref{sec:con}.

\section{The quantum Wasserstein distance of order 1 for qubits}\label{sec:W1}

The theory of optimal mass transport \cite{villani2008optimal,Ambrosio2008,ambrosio2013user} considers probability distributions on a metric space as distributions of a unit amount of mass.
The key element of such theory is the Monge--Kantorovich distance between probability distributions, which is the minimum cost that is required to transport one distribution onto the other, assuming that moving a unit of mass for a unit distance has cost one \cite{monge_wasser, kantorovich_wasser, vershik_2013}.
Such distance is also called earth mover's distance or Wasserstein distance of order $1$, often shortened to $W_1$ distance.
The exploration of the theory of optimal mass transport has led to the creation of an extremely fruitful field in mathematical analysis, with applications ranging from differential geometry and partial differential equations to machine learning \cite{Ambrosio2008, peyre2019computational,vershik_2013}.

The most natural distance on the set of the strings of $n$ bits is the Hamming distance~\cite{hamming_distance}, which counts the number of different bits.
The resulting $W_1$ distance on the set of the probability distributions on strings of $n$ bits is called Ornstein's $\bar{d}$-distance~\cite{ornstein1973application}.
Ref.~\cite{gdp_wasserstein_order_1} proposed a generalization of the $W_1$ distance to the space of the quantum states of a finite set of qubits, called quantum $W_1$ distance (or $QW_1$ in brief).
The generalization is based on the notion of neighboring quantum states.
Two quantum states of a finite set of qubits are neighboring if they coincide after discarding one qubit.
The quantum $W_1$ distance of Ref.~\cite{gdp_wasserstein_order_1} is the distance induced by the maximum norm that assigns distance at most $1$ to any couple of neighboring states; in the case of quantum states diagonal in the computational basis it recovers Ornstein's $\bar{d}$-distance
 and inherits most of its properties.

The $QW_1$ quantity can be computed with a semidefinite program, whose formulation requires to define a notion of Lipschitz constant for quantum observables.
The Lipschitz constant of the observable $\hat{H}$ acting on the Hilbert space of $n$ qubits is \cite{gdp_wasserstein_order_1}
\begin{equation}\label{eq:L}
\|\hat{H}\|_L = 2\max_{i=1,\,\ldots,\,n}\min_{\hat{H}_{i^c}}\left\|\hat{H} - \hat{\mathbb{I}}_i\otimes \hat{H}_{i^c}\right\|_\infty\,,
\end{equation}
where the minimization is performed over all the observables $\hat{H}_{i^c}$ that \emph{do not} act on the $i$-th qubit.
The quantum $W_1$ distance between the quantum states $\hat{\rho}$ and $\hat{\sigma}$ can then be expressed as \cite{gdp_wasserstein_order_1}
\begin{equation}
\|\hat{\rho} - \hat{\sigma}\|_{W_1} = \max_{\|\hat{H}\|_L\le1}\mathrm{Tr}\left[\left(\hat{\rho}-\hat{\sigma}\right)\hat{H}\right]\,.
\end{equation}
The present paper is based on the following lower bound to the quantum $W_1$ distance.
Let
\begin{equation}\label{eq:HW}
\hat{H}^{(\mathrm{wass})} = \sum_{i=1}^n |1\rangle_i\langle1| \otimes \hat{\mathbb{I}}_{i^c}\;,
\end{equation}
be the quantum observable that counts the number of ones in the computational basis.
We have $\left\|\hat{H}^{(\mathrm{wass})}\right\|_L = 1$ \cite{gdp_wasserstein_order_1}, therefore for any quantum state $\hat{\rho}$ we have
\begin{equation}
\left\|\hat{\rho} - |0\rangle\langle0|^{\otimes n}\right\|_{W_1} \ge \mathrm{Tr}\left[\hat{\rho}\,\hat{H}^{(\mathrm{wass})}\right]\,.
\end{equation}

$QW_1$ has found several applications in quantum information theory and many-body quantum physics, among which we mention a proof of the equivalence between the microcanonical and the canonical ensembles of quantum statistical mechanics \cite{de2022quantum} and a proof of limitations of VQA~\cite{de2022limitations,chou2022limitations}.
Furthermore, $QW_1$ has been extended to quantum spin systems on infinite lattices \cite{de2022wasserstein}.
In the context of quantum state tomography, the quantum $W_1$ distance has been employed as a quantifier of the quality of the learned quantum state and has led to efficient algorithms to learn Gibbs states of local quantum Hamiltonians \cite{rouze2021learning,maciejewski2021exploring,onorati2023efficient}.
In the context of quantum machine learning, the quantum $W_1$ distance has been employed as a cost function of the quantum version of generative adversarial networks \cite{gdp2021ghz,herr2021anomaly,anschuetz2022beyond,coyle2022machine}.

\subsection{Related approaches}

Several quantum generalizations of optimal transport distances have been proposed.
One line of research by Carlen, Maas, Datta and Rouz\'e \cite{carlen2014analog,carlen2017gradient,carlen2020non,rouze2019concentration,datta2020relating,van2020geometrical,wirth2022dual} defines a quantum Wasserstein distance of order $2$ from a Riemannian metric on the space of quantum states based on a quantum analog of a differential structure.
Exploiting their quantum differential structure, Refs. \cite{rouze2019concentration,carlen2020non,gao2020fisher} also define a quantum generalization of the Lipschitz constant and of the Wasserstein distance of order $1$.
Alternative definitions of quantum Wasserstein distances of order $1$ based on a quantum differential structure are proposed in Refs.~\cite{chen2017matricial,ryu2018vector,chen2018matrix,chen2018wasserstein}.
Refs.~\cite{agredo2013wasserstein,agredo2016exponential,ikeda2020foundation} propose quantum Wasserstein distances of order $1$ based on a distance between the vectors of the canonical basis.

Another line of research by Golse, Mouhot, Paul and Caglioti~\cite{golse2016mean,caglioti2021towards,golse2018quantum,golse2017schrodinger,golse2018wave, caglioti2019quantum,friedland2021quantum, cole2021quantum, duvenhage2021optimal,bistron2022monotonicity,van2022thermodynamic} arose in the context of the study of the semiclassical limit of quantum mechanics and defines a family of quantum Wasserstein distances of order $2$.
Ref.~\cite{de2021quantumAHP} proposes another quantum Wasserstein distance of order $2$ where the optimal transport is implemented with quantum channels.

The quantum Wasserstein distance between two quantum states can be defined as the classical Wasserstein distance between the probability distributions of the outcomes of an informationally complete measurement performed on the states, which is a measurement whose probability distribution completely determines the state.
This definition has been explored for Gaussian quantum systems with the heterodyne measurement
in Refs.~\cite{zyczkowski1998monge,zyczkowski2001monge,bengtsson2017geometry}.

\section{Preliminaries on QEC}\label{sec:introQEC}
Let $Q$ be a quantum register we wish to protect (at least in part) from the action of some external noise source.
In a typical QEC scenario~\cite{nielsen00} this problem is addressed through the following three-step procedure:
\begin{itemize}
 \item[{\it i)}] Before the action of the noise, a unitary encoding gate $\hat{V}_{QA}$ is used to distribute the information originally contained in $Q$ on the larger system
$QA$. Here $A$ is an auxiliary quantum register that is assumed to be initialized in a fiduciary quantum state, and that is affected by the same noise  that tampers with $Q$;
\item[{\it ii)}]
After the action of the noise, a measurement on $QA$ is performed to reveal the nature of the latter and,
based on the associated outcome, a unitary recovery operation is applied to the system. Equivalently
this step can be described by introducing yet an extra quantum register $B$ (also initialized on a fiduciary state but {\it not}
affected by the noise) that is coupled with $QA$ trough a
recovering unitary transformation $\hat{W}_{QAB}$ which effectively mimics the measurement and the recovery operation;
\item[{\it iii)}] The inverse of the gate $\hat{V}_{QA}$ is finally used on $QA$ to refocus the recovered information in $Q$.
\end{itemize}

Denoting with $|\psi\rangle_Q$ the input state of $Q$, the corresponding output state of $QA$ that emerges from the process at the end of the step {\it iii)} can be expressed as the density matrix
\begin{eqnarray}
&&\!\!\!\!\!\!\!\!\!\!\!\!\hat{\rho}^{(V,W)}_{QA}(\psi) := {\tr}_B\Big\{ {\cal V}^\dag_{QA}\circ {\cal W}_{QAB} \circ
\Phi_{Q A} \\
\nonumber
&&\quad \qquad \quad \circ {\cal V}_{QA} \Big(|\psi\rangle_{Q}\langle \psi| \otimes |\O\rangle_{A}\langle \O| \otimes |\O\rangle_{B}\langle \O|\Big)
\Big\} \\
&&\quad \quad := {\cal V}^\dag_{QA}\circ \Phi^{(R)}_{QA} \circ
\Phi_{Q A} \circ {\cal V}_{QA} \Big(|\psi\rangle_{Q}\langle \psi| \otimes |\O\rangle_{A}\langle \O| \Big) \nonumber
\end{eqnarray}
where $|\O\rangle_X$ represents the fiduciary state of the $X$ register, ${\tr}_B\{\cdots\}$ is the partial trace over $B$,
and given a unitary $\hat{U}_X$ on $X$ we adopted the symbol
 ${\cal U}_X(\cdots) := \hat{U}_X\cdots \hat{U}_X^\dag$ to denote its action as super-operator. In the above expressions $\Phi_{QA}$ is the LCPT
 quantum channel~\cite{nielsen00} describing the noise on $Q$ and $A$, while $ \Phi^{(R)}_{QA} (\cdots) : ={\tr}_B\{ {\cal W}_{QAB} (\cdots \otimes |\O\rangle_B\langle \O|)\}$ is the LCPT (recovery) quantum channel on $QA$ originating from the interaction with $B$, that attempts to undo the action of~$\Phi_{QA}$.

An ideal QEC procedure able to completely remove the noise from the system will make sure that $\hat{\rho}^{(V,W)}_{QA}(\psi)$ corresponds to $|\psi\rangle_{Q}|\O\rangle_A$,
 irrespectively from the specific choice of $|\psi\rangle_{Q}$.
A bona-fide figure of merit to characterize the effectiveness of a generic QEC scheme is hence provided
by the average input-output fidelity
\begin{eqnarray}\label{defFAV}
\overline{F}{(V,W)}&:=& \int d\mu_{\psi}\; {_{Q}\langle} \psi | {_{A}\langle} \O| \hat{\rho}^{(V,W)}_{QA}(\psi) |\psi\rangle_Q|\O\rangle_A \;,
\end{eqnarray}
where $d\mu_{\psi}$ is the uniform measure on the set of the input states of $Q$ originated
from the Haar measure on the associated unitary group~\cite{vinberg_groups_representations} or
from an exact or approximate unitary 2-design ${\cal S}$~\cite{2design_definition,nielsen00}
that simulates the latter\footnote{We remind that a unitary $2$-design is a probability distribution over the set of unitary operators which can duplicate properties of the probability distribution over the Haar measure for polynomials of degree $2$ or less. When $Q$ is a single qubit, a 2-design can be realized by a uniform sampling over a set ${\cal S}$ composed by only 6 elements $\iid$, $\hat{\sigma}_1$, $e^{\pm i \pi/4 \hat{\sigma}_1}$, $e^{\pm i \pi/4 \hat{\sigma}_2}$ that maps its logical state $|0\rangle_Q$ into the vectors $\ket{0}_Q, \ket{1}_Q, (\ket{0}_Q\pm i \ket{1}_Q)/\sqrt{2}, (\ket{0}_Q\mp \ket{1}_Q)/\sqrt{2}$.}.
Notice that by expressing $|\psi\rangle_Q= \hat{U}_Q |\O\rangle_Q$, Eq.~(\ref{defFAV}) can equivalently be casted in the more compact form
 \begin{eqnarray}\label{defFAV1}
\overline{F}{(V,W)}&=& {_{QA}\langle}\O| \hat{\rho}^{(V,W)}_{QA} |\O\rangle_{QA}\;,
\end{eqnarray}
with $|\O\rangle_{QA}:=|\O\rangle_Q\otimes |\O\rangle_A$ and where the state
  \begin{eqnarray}
\hat{\rho}^{(V,W)}_{QA} &:=&\frac{1}{|{\cal S}|} \sum_{\hat{U}_Q\in {\cal S}} \; {\cal U}^\dag_{Q} \circ {\cal V}^\dag_{QA}\circ \Phi^{(R)}_{QA} \circ
\Phi_{Q A} \nonumber \\
&&\circ\; {\cal V}_{QA}
 \circ \; {\cal U}_{Q} \Big(|\O\rangle_{QA}\langle \O| \Big)\;, \label{SAMP}
 \end{eqnarray}
 now includes the average over all possible inputs.
 An ideal QEC procedure will enable one to get $\overline{F}{(V,W)}=1$.
A natural benchmark for lowest admissible $\overline{F}{(V,W)}$ is represented instead by the value one would get if one decides not to perform corrections on the register that we compute by
 setting $\hat{V}_{QA}$ and $\hat{W}_{QAB}$ equal to the
identity operators
 i.e.\footnote{Equation~(\ref{fdffs}) accounts for
the noise effects both on $Q$ {\it and} $A$.
A more conservative estimation of $\overline{F}_0$ can be obtained by focusing directly on the noise on $Q$ alone, i.e.
tracing out the $A$ component of $\hat{\rho}^{(\openone,\openone)}_{QA}$ and studying its fidelity with $|\O\rangle_Q$, i.e.
 $\overline{F}^{(\rm strong)}_0 :=  {_{Q}\langle}\O| \hat{\rho}^{(\openone,\openone)}_{Q} |\O\rangle_{Q}\geq \overline{F}_0$,
with $\hat{\rho}^{(\openone,\openone)}_{Q}:=\tr_A\hat{\rho}^{(\openone,\openone)}_{QA}$. Notice that for the noise model
of Sec.~\ref{sec:noise} the two are directly connected via the identity
$\overline{F}_0= \overline{F}^{(\rm strong)}_0 - \frac{n-1}{n} p (1-|\langle 0|\hat{\sigma}|0\rangle|^2)$.}
  \begin{eqnarray} \label{fdffs}
 \overline{F}_0 :=  {_{QA}\langle}\O| \hat{\rho}^{(\openone,\openone)}_{QA} |\O\rangle_{QA} \;.
  \end{eqnarray}

 \begin{figure}[t]
\centering
		\includegraphics[width=\linewidth]{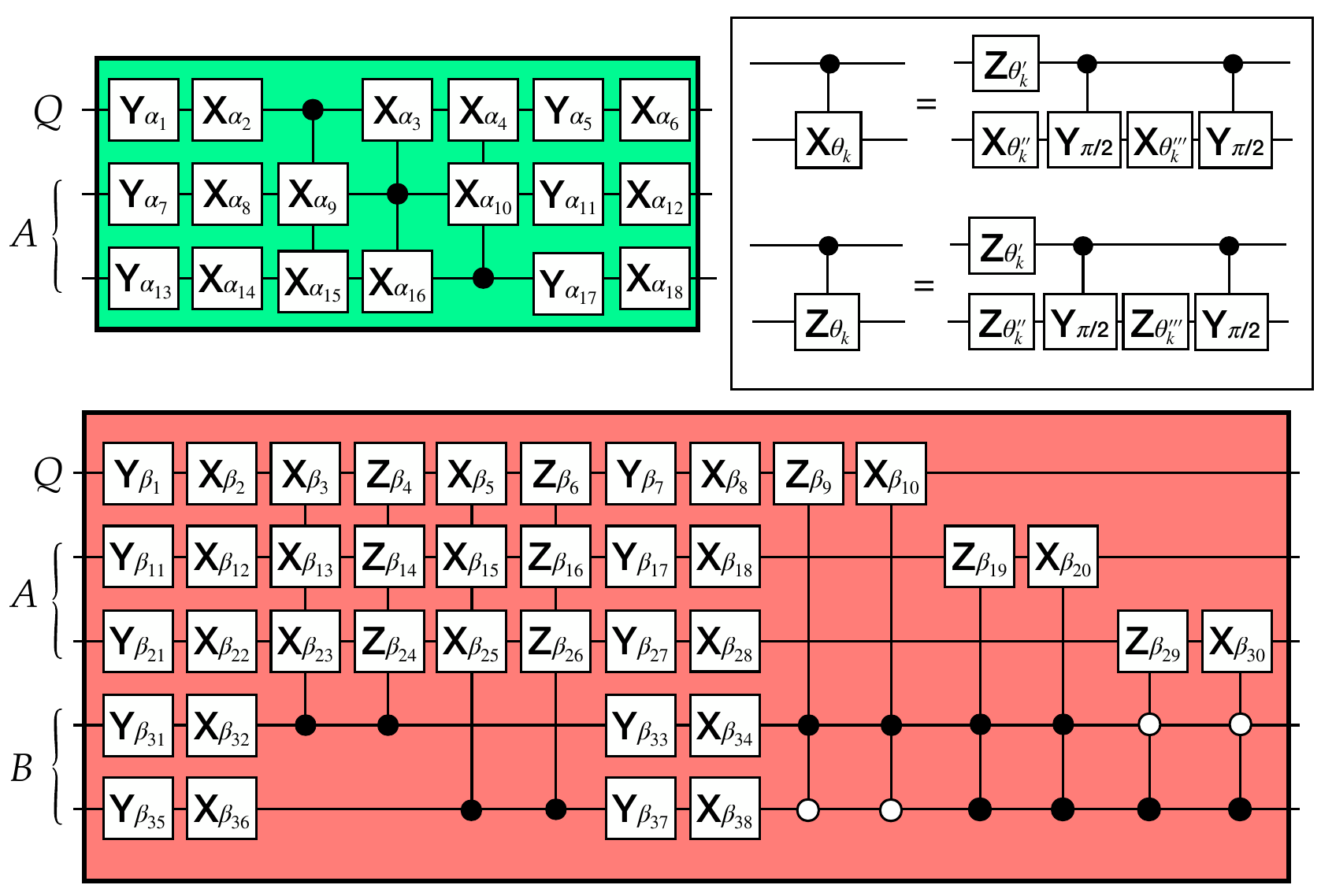}
		 \caption{Parametric gates $\hat{V}_{QA}(\vec \alpha)$ (green element) and $\hat{W}_{QAB}(\vec \beta)$
		(red element) used for case of quantum registers $Q$, $A$, and $B$ with $k=1$, $n-k=2$, and $r=2$ qubits respectively. Indicating with $\hat{\sigma}_1$,
 $\hat{\sigma}_2$, and $\hat{\sigma}_3$ the Pauli operators, the
 $X_{\theta}$, $Y_{\theta}$, and $Z_{\theta}$ elements of the figure represent single qubit rotations $e^{-i \theta \hat{\sigma}_1}$,
 $e^{-i \theta \hat{\sigma}_2}$, and $e^{-i \theta \hat{\sigma}_3}$
 with the angles $\theta$ determined to the components of the vectors
 $\vec{\alpha}$, $\vec{\beta}$, respectively. Vertical lines indicate instead quantum control operations which are activated when the control qubits (indicated by the full or empty circles) are
 in the logical state $|1\rangle$ (full circle) or in $|0\rangle$ (empty circle). As shown on the inset, each one of those
 gates depend parametrically upon elements of the control vectors $\vec{\alpha}$ and $\vec{\beta}$ through
 single qubit operations.
		 }\label{fig:solution_3qubit_V}
	\end{figure}

	\begin{figure}[]
		\centering
		\includegraphics[width=\linewidth]{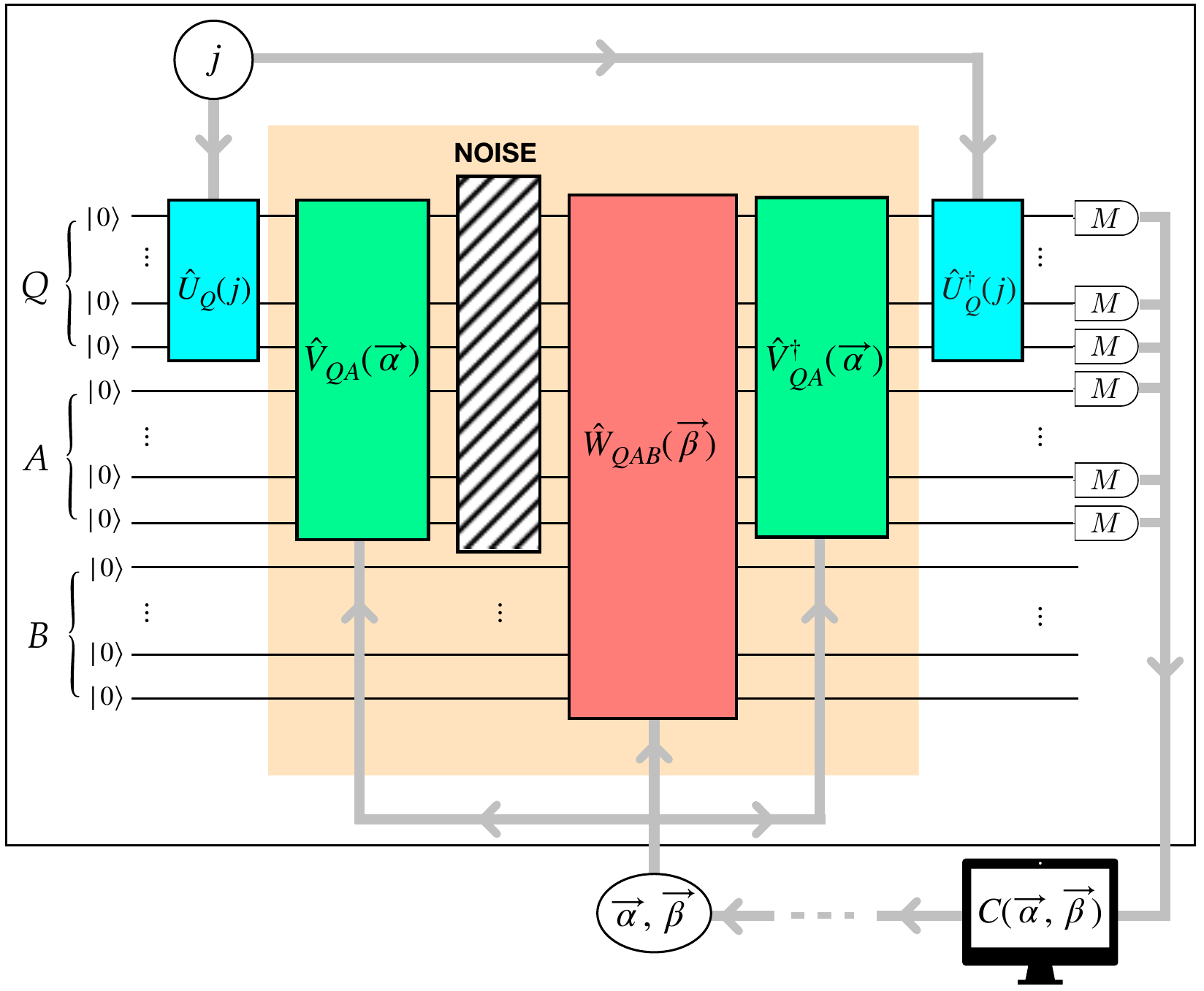}
		 \caption{(Color online) Sketch of the variational quantum algorithm: $Q$, $A$ and $B$ are quantum registers
		 formed respectively by $k$, $n-k$ and $r$ qubits.
		 The initial information we wish to protect is written in $Q$ by the unitary gate $\hat{U}_Q(j)$ extracted from a 2-design set ${\cal S}$;
		 $A$ and $B$ are two auxiliary elements (containing respectively $n-k$ and $r$ qubits) that are used to implement the QEC procedure described by the parametric gates $\hat{V}_{QA}(\vec{\alpha})$,
		 $\hat{W}_{QAB}(\vec{\beta})$, and $\hat{V}_{QA}^\dag(\vec{\alpha})$ of Fig.~\ref{fig:solution_3qubit_V}. The patterned element in the central part of the scheme represents the noise on $Q$ and $A$ (no noise is assumed to be active on $B$).
		 Lastly, the D-shaped measurements at the end of the circuit represent local measurements on $QA$ whose outcomes over
		 the entire collection of the possible inputs generated by the entire set ${\cal S}$,
		 are processed by a classical computer which, evaluating the cost function $C(\vec{\alpha},\vec{\beta})$ defined in \cref{sec:descent_algorithm},
		  decides how to update the values of the parameters $\vec{\alpha}$ and $\vec{\beta}$.
		  Thick grey lines in the figure represent classical control lines.}\label{fig:general_scheme_qcircuit}
	\end{figure}

\section{Variational Quantum Algorithm}\label{sec:general}
\begin{figure}[t]
\centering
		\includegraphics[width=\linewidth]{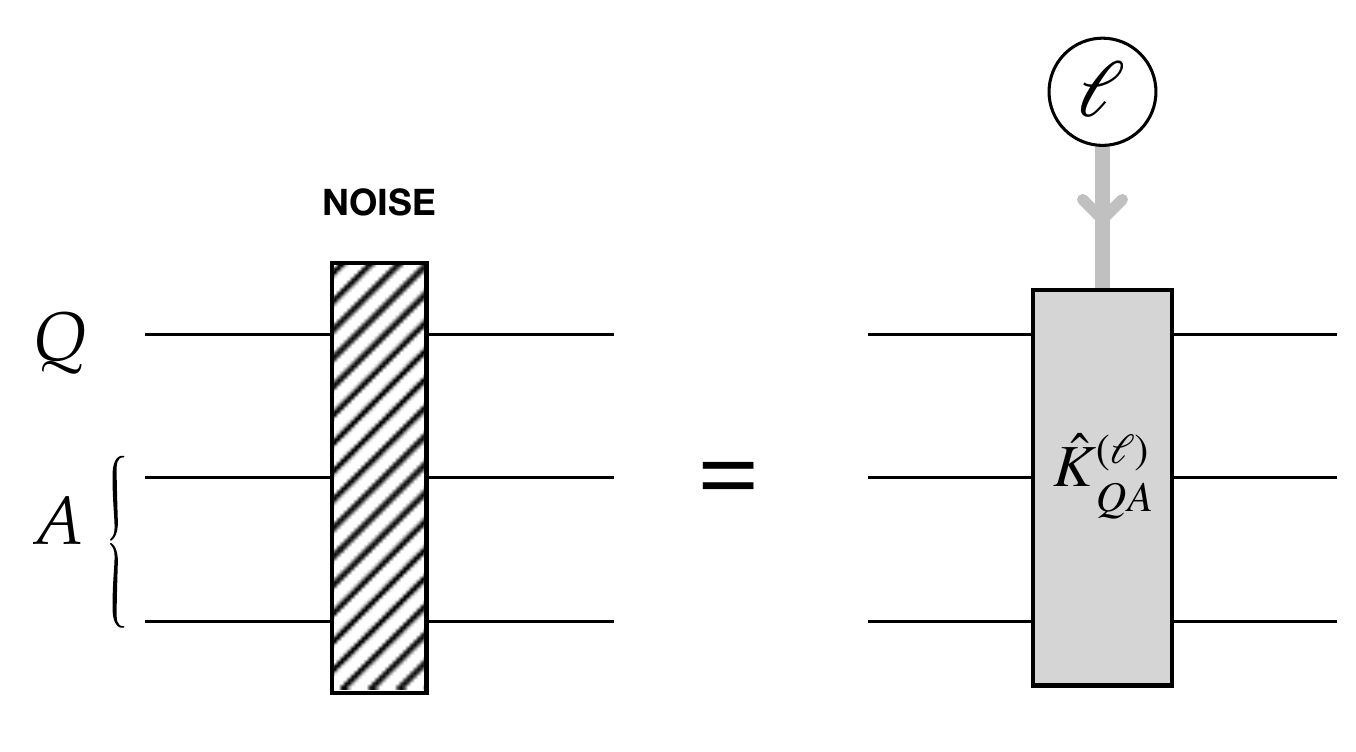}
		 \caption{Circuital implementation of the noise element of Fig.~\ref{fig:general_scheme_qcircuit}: here $\hat{K}_{QA}^{(\ell)}$
		 are weighted unitaries of Eq.~(\ref{kraus}).
		 	 }\label{fignoise}
	\end{figure}

While enormous progress has been made in the study of QEC procedures,  identifying
 efficient choices for the operations that lead to (non trivial) high values of $\overline{F}{(V,W)}$ for a
 specific noise model, is still a challenging open problem. A possible solution, in this case, is to employ variational quantum algorithms to run numerical searches.
Our approach follows a training strategy
 inspired by the work of Johnson \emph{et al.}~\cite{johnson2017qvector}.
 Assuming hence $Q$, $A$, and $B$ to be formed by collections of independent qubits ($k$ for
 $Q$, $n-k$ for $A$, and $r$ for $B$), we introduce a  manifold of transformations $\hat{V}_{QA}(\vec{\alpha})$, $\hat{W}_{QAB}(\vec{\beta})$ parametrized by  classical controls vectors $\vec{\alpha}$, $\vec{\beta}$ (see \cref{fig:solution_3qubit_V}),
 and construct the quantum circuit of~\cref{fig:general_scheme_qcircuit}.
The method then proceeds along the following stages:

\begin{enumerate}
\item Having selected the values of $\vec{\alpha}$ and~$\vec{\beta}$,
the register  $Q$
 is prepared into a collection of known quantum state $\{|\psi{(1)}\rangle_Q, \cdots , |\psi{(m)}\rangle_Q\}$ operating on the vector $|\O\rangle_Q=|0\rangle^{\otimes k}$ through action of the control
 gates $\hat{U}_Q{(1)},\cdots, \hat{U}_Q{(m)}$ (first cyan element of the figure) which define the 2-design ${\cal S}$ entering in Eq.~(\ref{SAMP}).
Each of such inputs is hence evolved via a circuit (pale-orange area of the figure)
 that emulates both the effect of the noise (patterned square of the figure, see~\cref{sec:noise} and Fig.~\ref{fignoise}), and the transformations $\hat{V}_{QA}(\vec{\alpha})$,
 $\hat{W}_{QAB}(\vec{\beta})$, and $\hat{V}_{QA}^\dag(\vec{\alpha})$ that are meant to implement the steps {\it ii)} and {\it iii)} of the
 QEC procedure (green and red elements of the figure). Notice that in the ideal case (i.e. if $\hat{V}_{QA}(\vec{\alpha})$ and $\hat{W}_{QAB}(\vec{\beta})$ manage to completely suppress the noise)
		then in correspondence with the input $|\psi{(j)}\rangle_Q$ the registers $QA$
		 should emerge in the state $|\psi{(j)}\rangle_Q\otimes |\O\rangle_A :=|\psi{(j)}\rangle_Q\otimes |0\rangle^{\otimes n-k}$, which
		 will be hence mapped into the final configuration $|\O\rangle_{QA}:=|0\rangle^{\otimes n}$
by the inverse $\hat{U}_Q^\dag(j)$ of the state preparation gate (second cyan element of the figure).

		 \item For each choice of the index $j\in\{1,\cdots,m\}$ a measurement on the system is performed at the end of the transformations described in stage 1 and the resulting $m$ collected outcomes
used to compute a cost function $C(\vec{\alpha},\vec{\beta})$ which evaluates the effectiveness of the adopted QEC strategy
in leading large values of the average input-output fidelity. The specific choice of the cost function is very important and is discussed in \cref{sec:cost_function}.

\item A classical computer decides, given the results of the measurement, how to change the value of the parameters
$\vec{\alpha}$ and $\vec{\beta}$ to be used in the subsequent run in order to minimize the cost function $C(\vec{\alpha},\vec{\beta})$. This is discussed in detail in \cref{sec:descent_algorithm}.
 \end{enumerate}

\subsection{Cost function}\label{sec:cost_function}
\begin{figure}[t]
\centering
		\includegraphics[width=\linewidth]{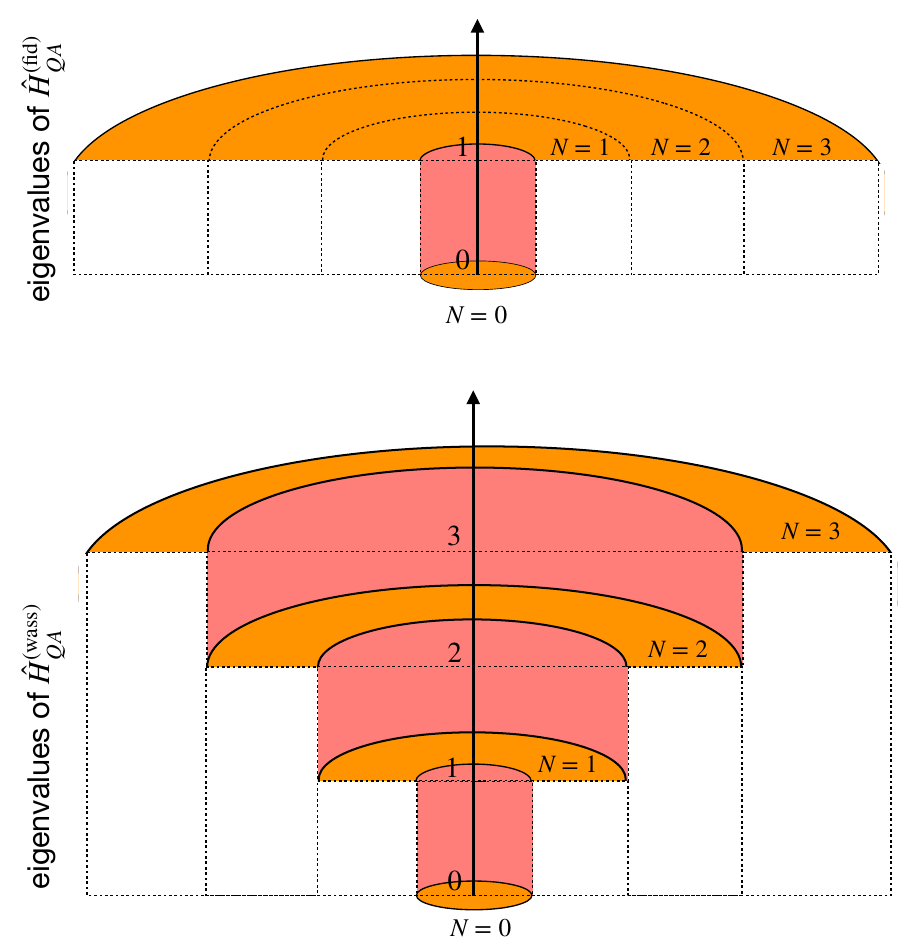}
		 \caption{Pictorial rendering of the spectra of the Hamiltonians $\hat{H}_{QA}^{(\rm fid)}$ (top panel) and
		 $\hat{H}_{QA}^{(\rm wass)}$ (lower panel). While $\hat{H}_{QA}^{(\rm fid)}$ is characterized by a unique, flat plateau
		that includes all the excited state, $\hat{H}_{QA}^{(\rm wass)}$ partially removes the associated degeneracy assigning
		higher energy to subspaces that have higher number of qubits in the logical state $|1\rangle$.
		 	 }\label{figurespectra}
	\end{figure}

The natural choice for the cost function at the stage 2 of our algorithm is provided by the expectation value of the self-adjoint
operator
\begin{eqnarray}\label{Hfid} \hat{H}^{(\rm fid)}_{QA}:= \iid_{QA} - \ket{\O}_{QA}\bra{\O}\;,\end{eqnarray} computed on the mean state of system $QA$ which emerges at the output of the quantum circuit of~\cref{fig:general_scheme_qcircuit}, i.e. the quantity
\begin{eqnarray} \label{costfid} C^{(\rm fid)}(\vec \alpha, \vec \beta) &:=&
\tr\{\rrho_{QA}^{(V(\vec{\alpha}),W(\vec{\beta}))} \hat{H}^{(\rm fid)}_{QA}\} \;,
\end{eqnarray}
where $\rrho_{QA}^{(V(\vec{\alpha}),W(\vec{\beta}))}$ is the density matrix (\ref{SAMP}) evaluated for
$\hat{V}_{QA}= \hat{V}_{QA}(\vec{\alpha})$ and $\hat{W}_{QAB}= \hat{W}_{QAB}(\vec{\beta})$.
This choice has two main advantages. First of all, the expectation value $C^{(\rm fid)}(\vec \alpha, \vec \beta)$ can be evaluated by
performing (simple) local measurement on the qubits of $Q$ and $A$ (indeed it can be computed by simply checking whether of not each
one of them is in the logical state $|0\rangle$). Most importantly, since
by explicit evaluation one has that $C^{(\rm fid)}(\vec \alpha, \vec \beta) = 1 - \overline{F}{(V(\vec{\alpha}),W(\vec{\beta})})$,
it is clear that by using (\ref{costfid}) the algorithm will be forced to look for values of $\vec \alpha$, $\vec \beta$ that yield
higher average input-output fidelities.
Despite all this, the use of $C^{(\rm fid)}(\vec \alpha, \vec \beta)$ as a cost function, has a major drawback associated with the fact that
the spectrum of the Hamiltonian $\hat{H}^{(\rm fid)}_{QA}$ exhibits maximum degeneracy with respect to space
orthogonal to the target state $|\O\rangle_{QA}$ (see Fig.~\ref{figurespectra}). Due to this fact
a numerical search based
on a training procedure that simply target the minimization of $C^{(\rm fid)}(\vec \alpha, \vec \beta)$, has non trivial chances to
 get stuck somewhere in the large flat plateau
associated with the eigenvalue 1 of $\hat{H}^{(\rm fid)}_{QA}$ without finding any good direction.
 in the large flat plateau
A possible way to avoid this problem is to introduce new cost-functions Hamiltonians which,
while maintaining the target vector $|\O\rangle_{QA}$ as a unique ground state and still being easy to compute, manage to remove the huge degeneracy of the excited part
of the spectra of $\hat{H}^{(\rm fid)}_{QA}$.
Our choice is based on the quantum Wasserstein distance of order 1 ($W_1$) introduced Ref.~\cite{gdp_wasserstein_order_1} which, even though it lacks some interesting properties that the fidelity has, is less likely to be affected by the barren plateaus phenomena~\cite{cerezo_barren_2021}.
As mentioned in Sec.~\ref{sec:W1}
 good estimation of the $W_1$ distance that separate $\rrho_{QA}^{(V(\vec{\alpha}),W(\vec{\beta}))}$ from the target state,
is provided by the following quantity
\begin{eqnarray}
 C^{(\rm wass)} (\vec \alpha, \vec \beta)   &:=& \tr\{\rrho_{QA}^{(V(\vec{\alpha}),W(\vec{\beta}))} \hat{H}_{QA}^{(\rm wass)} \}\;, \label{eq:cost_function} \\
\label{Hwass}
 \hat{H}_{QA}^{(\rm wass)} &:=&
 \dsum_{j = 1}^n j \; \hat{\Pi}_{QA}^{(j)} \;,
\end{eqnarray}
where $\hat{H}_{QA}^{(\rm wass)}$ is the Hamiltonian~(\ref{eq:HW}) which we express here
in terms of the projectors
$\hat{\Pi}^{(j)}_{QA}$ on  the sub-space of the register $QA$ in which we have $j$ qubits in $|1\rangle$ and the remaining one in $|0\rangle$.
Observe that, as already anticipated,
 $\hat{H}_{QA}^{(\rm wass)}$ is the sum of the number operators acting on the individual qubits of the register $QA$ as in \eqref{eq:HW}: accordingly, as $C^{(\rm fid)} (\vec \alpha, \vec \beta)$, $C^{(\rm wass)} (\vec \alpha, \vec \beta)$ can be computed from local measurement. What $C^{(\rm wass)} (\vec \alpha, \vec \beta)$ does is to count the total number of logical ones present in the system.
 To understand why using~(\ref{eq:cost_function}) could in principle lead to a more efficient numerical search than the one obtained by using (\ref{costfid}), notice that Eq.~(\ref{Hfid}) can be equivalently written as
$\hat{H}_{QA}^{(\rm fid)}
= \dsum_{j = 1}^n \hat{\Pi}^{(j)}_{QA}$. A comparison with~(\ref{Hwass}) reveals hence that indeed while both $\hat{H}_{QA}^{(\rm fid)}$ and
$\hat{H}_{QA}^{(\rm wass)}$ admit $|\O\rangle_{QA}$ as a unique ground state, the Wasserstein Hamiltonian removes large part of the degeneracy of the high energy spectrum of the fidelity Hamiltonian. Accordingly, it is reasonable to expect that a numerical search that uses $\hat{H}_{QA}^{(\rm wass)}$, has fewer chances to get trapped into regions of constant energy (barren plateau) than a search
based on  $\hat{H}_{QA}^{(\rm fid)}$,\footnote{It goes without mentioning that alternative choices for the cost function Hamiltonians are also available. For instance, one can use operators that also remove the
residual degeneracies that affect $\hat{H}_{QA}^{(\rm wass)}$ -- e.g. using the operator
$\hat{H}_{QA}^{(\rm full)} = \sum_{\ell =1}^n w_{\ell} \hat{\pi}_\ell$ with $\omega_\ell$ positive weights selected so that
different allocation of $|1\rangle$ states inside the eigenspaces of $\hat{H}_{QA}^{(\rm wass)}$ get an assigned ordering.
Our numerical analysis however seems to indicate that these refinements do not contribute significantly in improving numerical search of the algorithm.}.

\subsection{Descent algorithm}\label{sec:descent_algorithm}
The algorithm that we used for this work is a gradient descent algorithm with momentum~\cite{NoceWrig06}. To overcome the numerical difficulties of using finite differences to estimate the gradients of the cost function $C(\vec{\alpha},\vec{\beta})$, we exploit a variation of the parameter-shift rule introduced in~\cite{schuld_analytical_gradients} which reduces the problem to compute linear combinations of the function itself evaluated in different points that are not infinitesimally close.
 Specifically, we observe
that, irrespectively from the choice of the operator $\hat{H}_{QA}$, the functional dependence of $C(\vec \alpha, \vec \beta)$ upon the $j$-th component of the vector $\vec{\beta}$
is of the form
\begin{equation}
 \label{eq:example_parameter_shift}
 C(\vec \alpha, \vec \beta) = f(\beta_j):=\sum_{k} \tr\big\{\hat{\Omega}^{(k)}_1e^{i\beta_j \hat{\sigma}}\hat{\Omega}^{(k)}_2 e^{-i\beta_j\hat{\sigma}}\big\},
\end{equation}
with $\hat{\Omega}_{1,2}^{(k)}$ being multi-qubits operators which do not depend upon $\beta_j$, and with $e^{-i\beta_j \hat{\sigma}}$ a single qubit rotation generated by an element $\hat{\sigma}$ of the Pauli set. Therefore its gradient can be written as
\begin{eqnarray}
 \label{eq:gradient_example_parameter_shift}
 \frac{\partial C(\vec \alpha, \vec \beta) }{\partial \beta_j} &=& i \sum_k \tr\big\{ \hat{\Omega}^{(k)}_1e^{i\beta_j \hat{\sigma}} [\hat{\sigma},\hat{\Omega}^{(k)}_2] e^{-i\beta_j \hat{\sigma}}\big\} \nonumber \\
 &=& f(\beta_j + \tfrac{\pi}{4}) - f(\beta_j - \tfrac{\pi}{4})\;,
\end{eqnarray}
where in the last passage we used the identity
\begin{align}\label{eed}
 i [\hat{\sigma},\hat{\Omega}^{(k)}_2]= e^{i\frac{\pi}{4} \hat{\sigma}} \hat{\Omega_2}^{(k)} e^{-i\frac{\pi}{4}\hat{\sigma}} - e^{-i\frac{\pi}{4} \hat{\sigma}} \hat{\Omega_2}^{(k)} e^{i\frac{\pi}{4}\hat{\sigma}}.
\end{align}
The gradient with respect the vector $\vec{\alpha}$ can be computed similarly. In this case however we observe that, due to the fact that $\hat{\rho}^{(V(\vec{\alpha}),W(\vec{\beta}))}_{QA}(\psi)$ depends upon
the parameters $\vec{\alpha}$ via $\hat{V}_{QA}(\vec{\alpha})$ and through its
adjoint $\hat{V}_{QA}^\dag(\vec{\alpha})$,
 the dependence of $C(\vec \alpha, \vec \beta)$ upon the $j$-th component of $\vec{\alpha}$
is slightly more complex. Indeed in this case we have
\begin{eqnarray}
 \label{eq:example_parameter_shift_alpha}
 C(\vec \alpha, \vec \beta) &=& g(\alpha_j,\alpha_j) \;, \end{eqnarray}
  where $g(\alpha^{(1)}_j,\alpha^{(2)}_j)$ is the function
  \begin{eqnarray}
g(\alpha^{(1)}_j,\alpha^{(2)}_j)
 :=\sum_k  &&\tr\big\{\hat{\Omega}^{(k)}_1e^{i\alpha^{(1)}_j \hat{\sigma}}\hat{\Omega}^{(k)}_2 e^{-i\alpha^{(1)}_j\hat{\sigma}} \\\nonumber
 && \qquad \times \hat{\Omega}^{(k)}_3 e^{i\alpha^{(2)}_j\hat{\sigma}}
 \hat{\Omega}^{(k)}_4 e^{-i\alpha^{(2)}_j\hat{\sigma}}
 \big\}\;,
\end{eqnarray}
with $\hat{\Omega}^{(k)}_{1,2,3,4}$ representing multi-qubits operators which do not depend neither upon
 $\alpha_j^{(1)}$ nor $\alpha_{j}^{(2)}$. It is important to stress that $g(\alpha^{(1)}_j,\alpha^{(2)}_j)$ can be computed using the same circuit
 of Fig.~\ref{fig:general_scheme_qcircuit}, by simply replacing the phases $\alpha_j$ of $\hat{V}_{QA}(\vec{\alpha})$ and $\hat{V}_{QA}^\dag(\vec{\alpha})$ with
 $\alpha_j^{(1)}$ and $\alpha_j^{(2)}$ respectively.
Notice finally that exploiting the identity Eq.~(\ref{eed}) we can write
 \begin{eqnarray}
 \label{eq:gradient_example_parameter_shift_alpha}
 \frac{\partial C(\vec \alpha, \vec \beta) }{\partial \alpha_j} &=& \left. \frac{\partial g(\alpha^{(1)}_j,\alpha_j)}{\partial \alpha^{(1)}_j}\right|_{\alpha_j^{(1)}=\alpha_j}
 + \left.\frac{\partial g(\alpha_j,\alpha^{(2)}_j)}{\partial \alpha^{(2)}_j}\right|_{\alpha_j^{(2)}=\alpha_j}
 \\
 &=& g(\alpha_j + \tfrac{\pi}{4},\alpha_j) - g(\alpha_j -\tfrac{\pi}{4},\alpha_j) \nonumber \\
 &+& g(\alpha_j,\alpha_j + \tfrac{\pi}{4}) - g(\alpha_j,\alpha_j -\tfrac{\pi}{4})\nonumber\;,
\end{eqnarray}
which shows that computing the gradient of $C(\vec \alpha, \vec \beta)$ with respect to $\alpha_j$ simply accounts to evaluate
the circuit that express $g(\alpha_j^{(1)},\alpha_j^{(2)})$ for four distinct values of the parameters.

\subsection{Noise model}\label{sec:noise}
\begin{figure}[t]
\centering
		\includegraphics[width=\linewidth]{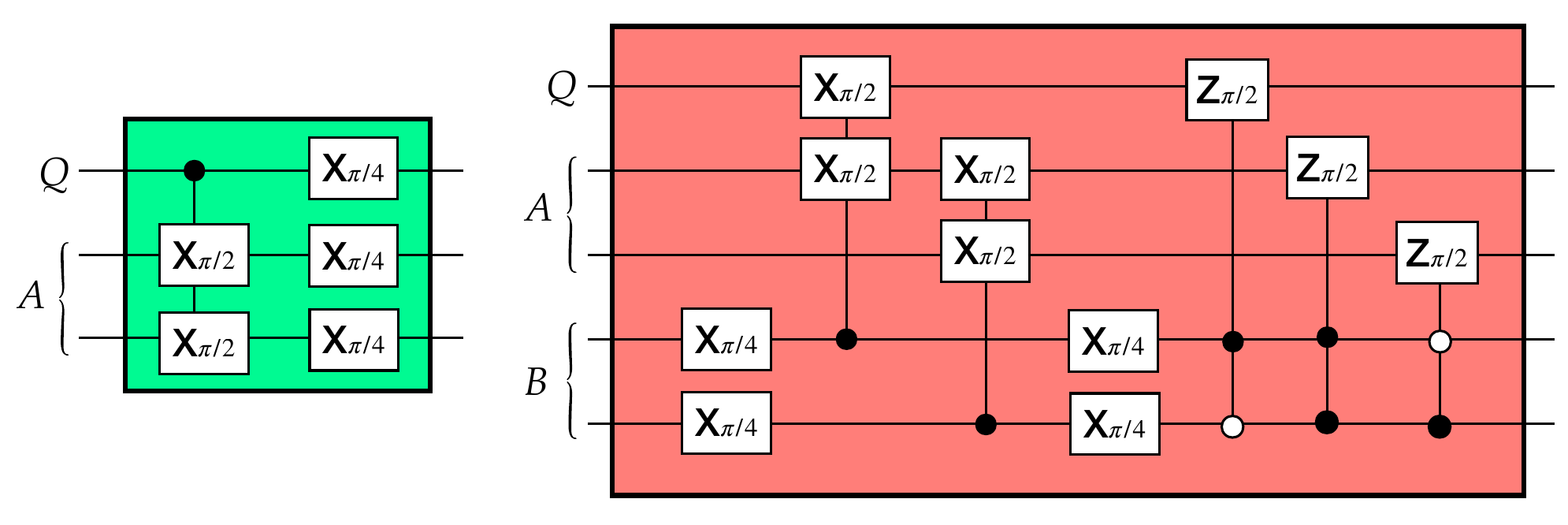}
		 \caption{Circuital implementations of the ideal transformations
		 $\hat{V}_{QA}(\vec \alpha)$ (left) and $\hat{W}_{QAB}(\vec \beta)$ (right) which allow for exact noise suppression
		 of a single-qubit bit-flip noise model [i.e. (\ref{noise}) with $\hat{\sigma}^{(\ell)}=\hat{\sigma}^{(\ell)}_1$]
		 using a quantum register $B$ with $r=2$ qubit.
		 gates.
		 	 }\label{FIGexact}
	\end{figure}

The scheme presented so far can in principle be applied to arbitrary classes of noises.
In our research however we focused on a specific model that has been extensively studied in the literature producing explicit examples of efficient QEC solutions which can be used
as a theoretical benchmark for our variational search.
Specifically we assume
$Q$ and $A$ to be respectively a single qubit register ($k=1$) and a two qubit register ($n=3$), globally affected by a given species of single-qubit
noise~\cite{gottesman_2009,Knill_2001}.
These transformations can be represented in terms of a LCPT map of the form
\begin{eqnarray}
\label{noise}
\Phi_{QA}(\cdots) = \sum_{\ell=0}^{n} \hat{K}^{(\ell)}_{QA} \cdots \hat{K}^{(\ell)\dag}_{QA}\;,
\end{eqnarray}
with
Kraus operators~\cite{nielsen00}
\begin{align} \label{kraus}
 \hat{K}_{QA}^{(0)} := \sqrt{1 - p} \; \iid_{QA} \;, \qquad \hat{K}_{QA}^{(\ell)}:= \sqrt{\frac{p}{n}} \; \hat{\sigma}^{(\ell)}\;,
\end{align}
where for $\ell\in\{1,\cdots,n\}$, $\hat{\sigma}^{(\ell)}$ is the Pauli operator acting on the $\ell$-th qubit of $QA$ which defines the noise species we have selected. For instance, in the case we choose to describe phase-flip noise then $\hat{\sigma}^{(\ell)}=\hat{\sigma}^{(\ell)}_3$, while for describing bit-flip we have $\hat{\sigma}^{(\ell)}=\hat{\sigma}^{(\ell)}_1$.
Explicit examples of $\hat{V}_{QA}$, $\hat{W}_{QAB}$ which allow for exact suppression of the noise ($\overline{F}{(V,W)}=1$) are
shown in Fig.~\ref{FIGexact}. Notice that by construction the circuit parametrization of
 $\hat{V}_{QA}(\vec \alpha), \hat{W}_{QAB}(\vec \beta)$ given in Fig.~\ref{fig:general_scheme_qcircuit} include such gates
 as special solution: accordingly if properly guided by an efficient cost function, our numerical VQA search has a chance to
 find the solution of Fig.~\ref{FIGexact}.

\section{Results}\label{sec:res}

\begin{figure}[t]
 \includegraphics[width=\linewidth]{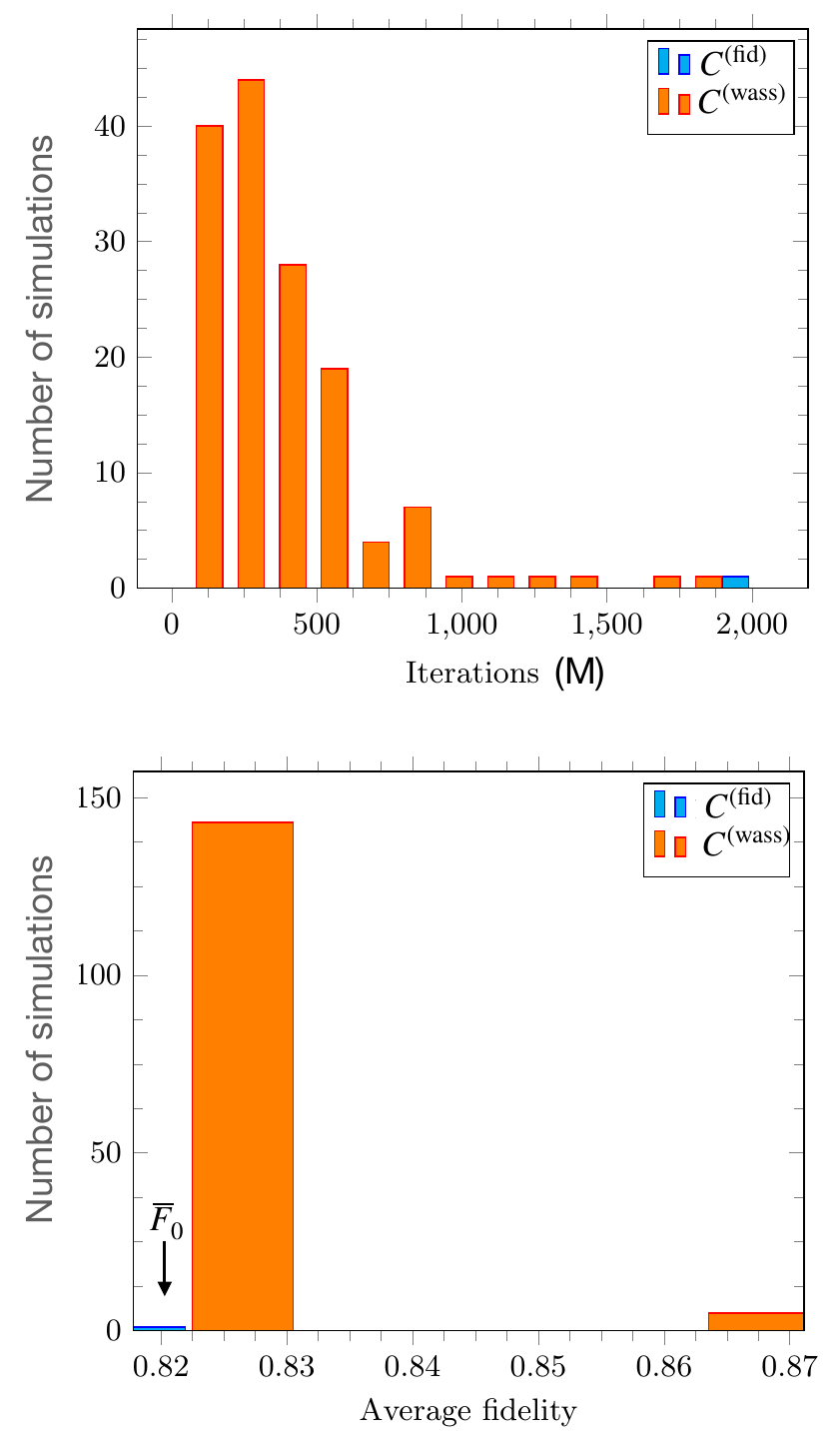}
 \caption{Comparison of the input-output average fidelity~(\ref{defFAV1})
  attainable by running our optimization algorithm
  using the cost function $C^{(\rm fid)}(\vec \alpha, \vec \beta)$ (blue data) and $C^{(\rm wass)}(\vec \alpha, \vec \beta)$ (orange data).
 Here the error model is a single-qubit bit-flip noise ($\hat{\sigma}=\hat{\sigma}_1$ in (\ref{noise})) with $p = 0.8$. The no error correction threshold (\ref{fdffs}) of this scheme is
$\overline{F}_0 \approx 0.822$ -- orange peak in the fidelity plot, up to numerical precision. Only the runs that produced a fidelity of at least $\overline{F}_0$ have been included: for $C^{(\rm fid)}(\vec \alpha, \vec \beta)$ this happens in the $0.2\%$ of the runs, while for $C^{(\rm wass)}(\vec \alpha, \vec \beta)$ for the $29.6\%$.}\label{fig:qvector_trick_results_sigmax}
\end{figure}

\begin{figure}[t]
\includegraphics[width=\linewidth]{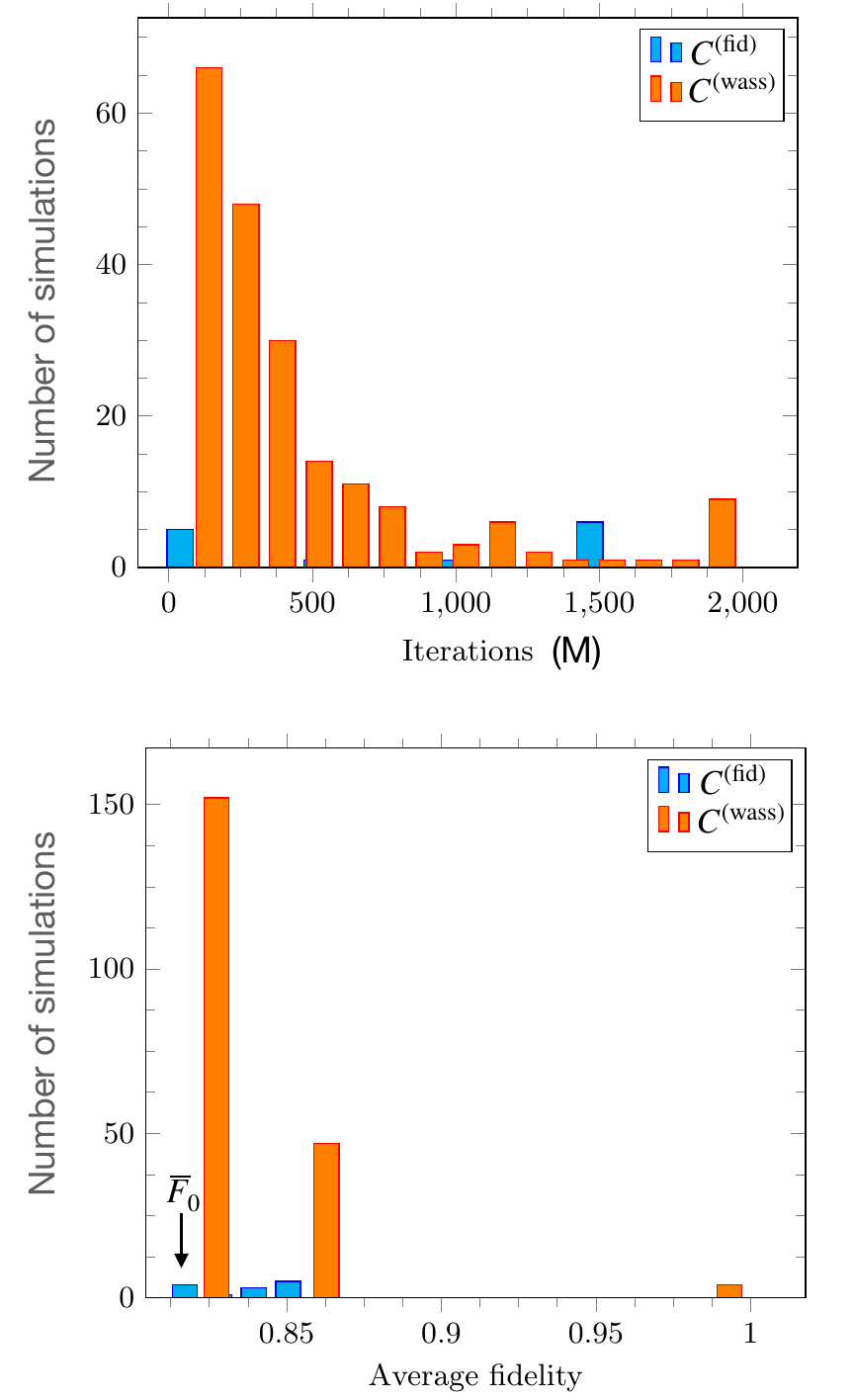}
 \caption{Comparison of the the input-output average fidelity~(\ref{defFAV1})
  attainable by running our optimization algorithm
  using the cost function $C^{(\rm fid)}(\vec \alpha, \vec \beta)$ (blue data) and $C^{(\rm wass)}(\vec \alpha, \vec \beta)$ (orange data).
 Here the error model is a single-qubit phase-flip noise ($\hat{\sigma}=\hat{\sigma}_3$ in (\ref{noise}) with $p = 0.8$. The no error correction threshold (\ref{fdffs}) of this scheme is
$\overline{F}_0 \approx 0.822$ -- orange peak in the fidelity plot, up to numerical precision. Only the runs that produced a fidelity of at least $\overline{F}_0$ have been included: for $C^{(\rm fid)}(\vec \alpha, \vec \beta)$ this corresponds to the $2.6\%$ of the runs, while for $C^{(\rm wass)}(\vec \alpha, \vec \beta)$ the success probability is $40.6\%$.}\label{fig:qvector_trick_results_sigmaz}
\end{figure}

In this section we study the impact of the cost function on the
 efficiency of the optimization algorithm of Sec.~\ref{sec:general}. Assuming the single-qubit noise model detailed in Sec.~\ref{sec:noise} and taking $B$ to be a $r=2$ qubit register, we run two distinct numerical searches:
 the first obtained by identifying $C(\vec \alpha, \vec \beta)$ with $C^{(\rm fid)}(\vec \alpha, \vec \beta)$ and the second choosing instead
  $C^{(\rm wass)}(\vec \alpha, \vec \beta)$. Results are reported in
 \cref{fig:qvector_trick_results_sigmaz,fig:qvector_trick_results_sigmax} for two different choices of the noise models~(\ref{noise}), i.e. phase-flip and bit-flip.
  For both, we compare the input-output average fidelity~(\ref{defFAV1}) at the end of the procedure obtained with the two
  different cost functions, and the number of iterations $M$ needed for convergence.
Regarding this last quantity we set a maximum value $M_{\max}$ equal to 2000 before convergence and we chose this limit mainly with practical choices like the maximum time for the simulation, enforcing that a single run does not require more than a few hours of computational time: in case the algorithm fails to reach the convergency we simply stop the numerical search (this is the reason for the peak at the end of the upper orange plot in \cref{fig:qvector_trick_results_sigmaz}).
The plots report only the simulations that
manage to achieve an average fidelity that is greater or equal than no-correction threshold bound $\overline{F}_0$.

The first thing to observe is that for both noise models,
$C^{(\rm fid)}(\vec \alpha, \vec \beta)$ has problem in reaching the do-nothing threshold $\overline{F}_0$: the probability of success
being $2.6\%$ for the
phase-flip case of~\cref{fig:qvector_trick_results_sigmaz} and only $0.2\%$ for the bit-flip case of~\cref{fig:qvector_trick_results_sigmax} (for both noise models the total number of simulations analyzed was 500).
Observe also that in this last case the algorithm never yields average input-output fidelity values strictly larger than $\overline{F}_0$ and that, even in those
cases, it requires a number $M$ of iterations which saturate the maximum allowed value $M_{\max}$ (blue peak in the upper plot of \cref{fig:qvector_trick_results_sigmaz}).
$C^{(\rm was)}(\vec \alpha, \vec \beta)$ performs definitely better: to begin with
it succeeds in overcoming the threshold $\overline{F}_0$ in one third of the simulations (specifically
 $40.6\%$ for the phase-flip noise model and $29.6\%$ for the bit-flip noise model). Furthermore, the algorithm reaches convergency
 with a number of iterations that are typically smaller than those required by $C^{(\rm fid)}(\vec \alpha, \vec \beta)$.

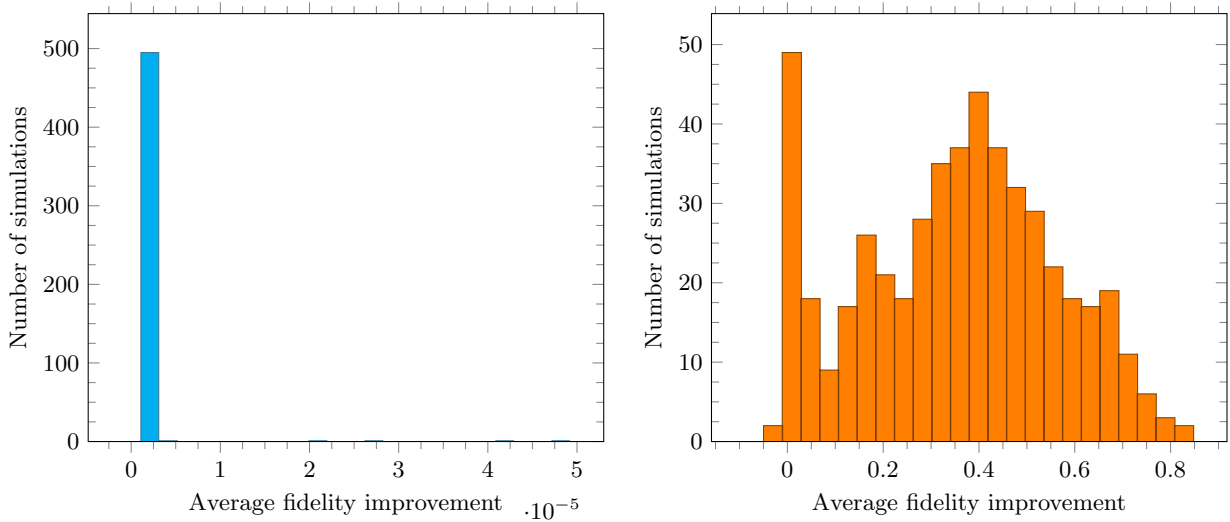
\begin{figure}[th]
 \centering
  \begin{tikzpicture}
  \begin{axis}[
   ymin=0,
   minor y tick num = 3,
   minor x tick num = 3,
   area style,
   xlabel={Average fidelity improvement},
   ylabel={Number of simulations},
   bar width=0.000002,
   ybar
   ]
   \addplot+[fill=cyan, draw=cyan!50!black] coordinates {
    % f0b76713d09ad49f26785bab75276a72
    (4.817572201687437e-05, 1) (4.608107650704441e-05, 0) (4.3986430997214455e-05, 0) (4.1891785487384496e-05, 1) (3.9797139977554537e-05, 0) (3.770249446772458e-05, 0) (3.5607848957894625e-05, 0) (3.3513203448064665e-05, 0) (3.1418557938234706e-05, 0) (2.9323912428404753e-05, 0) (2.7229266918574794e-05, 1) (2.5134621408744838e-05, 0) (2.3039975898914882e-05, 0) (2.0945330389084923e-05, 1) (1.8850684879254967e-05, 0) (1.675603936942501e-05, 0) (1.4661393859595051e-05, 0) (1.2566748349765092e-05, 0) (1.047210283993514e-05, 0) (8.37745733010518e-06, 0) (6.282811820275221e-06, 0) (4.188166310445268e-06, 1) (2.093520800615309e-06, 495) (1.1247092146504656e-09, 0)
   };
  \end{axis}
 \end{tikzpicture}
 \quad
 \begin{tikzpicture}
  \begin{axis}[
   ymin=0,
   minor y tick num = 3,
   minor x tick num = 3,
   area style,
   xlabel={Average fidelity improvement},
   ylabel={Number of simulations},
   bar width=0.04,
   ybar
   ]
   \addplot+[fill=orange, draw=orange!50!black] coordinates {
    % 7890a7676a1c37adefa6dd9bc5ae54b8
    (0.8280638490853384, 2) (0.789073049775518, 3) (0.7500822504656975, 6) (0.7110914511558771, 11) (0.6721006518460567, 19) (0.6331098525362363, 17) (0.594119053226416, 18) (0.5551282539165956, 22) (0.5161374546067752, 29) (0.4771466552969547, 32) (0.4381558559871343, 37) (0.3991650566773139, 44) (0.36017425736749353, 37) (0.3211834580576731, 35) (0.2821926587478527, 28) (0.2432018594380323, 18) (0.20421106012821189, 21) (0.16522026081839147, 26) (0.12622946150857106, 17) (0.08723866219875065, 9) (0.04824786288893024, 18) (0.009257063579109825, 49) (-0.029733735730710587, 2) (-0.068724535040531, 0)
   };
  \end{axis}
 \end{tikzpicture}
 \caption{Improvement of simulations when changing the cost function in a two run optimization process that uses
different cost functions to drive the descent algorithm. In the left plot, we started the descent on a random initial point, ran the optimization using $C^{(\rm wass)}(\vec \alpha, \vec \beta)$ as cost function until convergence, and then we started the descent algorithm again but using $C^{(\rm fid)}(\vec \alpha, \vec \beta)$ as cost function, starting from the final point of the previous descent. In the right part, the roles of the two cost functions are inverted (we start using $C^{(\rm fid)}(\vec \alpha, \vec \beta)$ and then we use $C^{(\rm wass)}(\vec \alpha, \vec \beta)$). The histograms represent the difference in average input-output fidelity~(\ref{defFAV}) after the change of cost function, namely the difference between the fidelity achieved after the second descent and the fidelity after the first descent (positive values correspond to
improved performances).Please notice the scale difference on the $x$-axis between the left and right plot.}\label{fig:performance_start_differed}
\end{figure}

 To better enlighten the differences between the two cost functions, we proceeded with further simulations, whose results are summarized in \cref{fig:performance_start_differed}.
The idea here is to run a two-step optimization process composed by two sequences of runs:
in the first run we start the optimization procedure from a random point in the parameter space
  $(\vec \alpha, \vec \beta)$  with one of the two cost functions (say $C^{(\rm fid)}(\vec \alpha, \vec \beta)$), up to convergence; after that we start a second optimization run using
  the other cost function (say $C^{(\rm wass)}(\vec \alpha, \vec \beta)$) but assuming as initial condition for the parameters the final point reached by the first run. The
  plots report the difference in fidelity between the second and the first run: when we start using the $C^{(\rm wass)}(\vec \alpha, \vec \beta)$ in the first run, the fidelity cannot further improve the result that is already found, and this is represented by the fact that the best improvement is of the order of $10^{-5}$; on the contrary, if we started employing $C^{(\rm fid)}(\vec \alpha, \vec \beta)$ in the first run, the use of
  $C^{(\rm wass)}(\vec \alpha, \vec \beta)$ in the second run typically yields substantial improvements of the performance\footnote{It has to be said that in few cases the figure of merit is worse after the second optimization -- see the negative bar in right panel of~\cref{fig:performance_start_differed}. This is due to the fact that when using $C^{(\rm wass)}(\vec \alpha, \vec \beta)$ we are not maximizing the fidelity but minimizing a function whose stationary point corresponds to the maximum of the latter: accordingly the final point of convergence for  $C^{(\rm wass)}(\vec \alpha, \vec \beta)$ can be slightly off mark in terms of fidelity. This is not a problem because these two functions do not have a constant ratio, and we checked that the inequalities between them are still satisfied.}. Moreover, we sampled some single descent processes and plotted the cost in function of the iteration. When we move from fidelity to $W_1$, the descent part after the change of cost function is qualitatively indistinguishable from starting from a random point.

\onecolumngrid%

\twocolumngrid%

\section{Conclusions}\label{sec:con}

To summarize, we have shown a variational quantum algorithm that allows finding the most suitable error correction procedure for a specific noise on quantum hardware. We compared the performance of two different versions of this algorithm using two different cost functions, the fidelity and an approximation of the quantum Wasserstein distance of order one. We compared the difference in speed and the ability to obtain a useful solution between the two algorithms, finding really different trends between the two optimization procedures. The optimization process based on the fidelity suffers greatly from the phenomenon of the barren plateaus, leading to very slow convergence or no convergence at all, while the algorithm based on the quantum $W_1$ distance allows us to find the configurations that correct the errors in the examples that we explored. The obtained results show a clear improvement and allow us to explore further improvements of these methods, as using different algorithms for the minimization process, e.g. stochastic gradient descent or higher-order algorithms like Newton or pseudo-Newton algorithms.

Given that the gradient can be expressed only with the cost function evaluated in a small number of circuits that differ only in the parameter choice, the gradient of the cost function can be computed on the same hardware that will be used for the correction procedure. Moreover, simulating this circuit may be difficult because of the exponential scaling of the dimension of the Hilbert space of a set of qubits, but this problem does not apply when all the circuit is built on hardware, gaining a quantum advantage.
For the same reason, the same procedure can be iterated to compute the exact Hessian of the cost function and then apply a second-order method like the Newton method as a descent algorithm. However, this has not been done because the circuits that we marked as useful have a relatively big number of parameters, and computing the hessian scales quadratically with this number, leading to intractable computations.

With this work, we have shown a clear advantage in the use of this $QW_1$ distance approximation in a gradient-based optimization algorithm. In future work, it may be interesting to study the effect of this cost function also on gradient-free optimization algorithms, that do also suffer from the barren-plateaus phenomenon, as shown in Ref.~\cite{cerezo_gradient_free}.

\subsection*{Acknowledgments}

FZ and VG acknowledge financial support by MIUR (Ministero dell’ Istruzione, dell’ Universit\`a della Ricerca) by PRIN 2017 Taming complexity via Quantum Strategies: a Hybrid Integrated Photonic approach (QUSHIP) Id. 2017SRN-BRK, and via project PRO3 Quantum Pathfinder. GDP is a member of the ``Gruppo Nazionale per la Fisica Matematica (GNFM)'' of the ``Istituto Nazionale di Alta Matematica ``Francesco Severi'' (INdAM)''.
GDP has been supported by the HPC National Centre for HPC, Big Data and Quantum Computing – Proposal code CN00000013, CUP J33C22001170001, funded within PNRR - Mission 4 - Component 2 Investment 1.4.\ SL was funded by ARO and by DARPA.
MM is supported by the NSF Grants No. CCF-1954960 and CCF-2237356.

\bibliography{bibliografia.bib}
%
% \appendix

\end{document}